# Highly efficient ultra-broad beam silicon nanophotonic antenna based on near-field phase engineering

Shahrzad Khajavi,[1,*] Daniele Melati,[2] Pavel Cheben,[3] Jens H. Schmid,[3] Carlos A. Alonso Ramos,[2] and Winnie N. Ye[1]

[1]Department of Electronics, Carleton University, 1125 Colonel By Drive, Ottawa, ON K1S 5B6, Canada
[2]Centre for Nanoscience and Nanotechnologies, CNRS, Université Paris-Saclay, 10 Bv. Thomas Gobert, 91120 Palaiseau, France
[3]Advanced Electronics and Photonics Research Center, National Research Council Canada, 1200 Montreal Road, Ottawa, ON K1A 0R6, Canada
[*]shahrzadkhajavi@cmail.carleton.ca

## Abstract

Optical antennas are a fundamental element in optical phased arrays (OPA) and free-space optical interconnects. An outstanding challenge in optical antenna design lies in achieving high radiation efficiency, ultra-compact footprint and broad radiation angle simultaneously, as required for dense 2D OPAs with a broad steering range. Here we demonstrate a fundamentally new concept of a nanophotonic antenna based on near-field phase-engineering. By introducing a specific near-field phase factor in the Fraunhofer transformation, the far-field beam is widened beyond the diffraction limit for a given aperture size. We use transversally interleaved subwavelength grating nanostructures to control the near-field phase. The antenna reaches a radiation efficiency of 82%, a compact footprint of 3.1 µm × 1.75 µm and an ultra-broad far-field beam width of 52° and 62° in the longitudinal and transversal direction, respectively. This unprecedented design performance is achieved with a single-etch grating nanostructure in a 300-nm SOI platform.

## 1. Introduction

Optical beam steering is a fundamental functionality at the core of many technologies, including free-space optical communications, 3D imaging and mapping, interconnects and optical memories [1–3]. Ideally, a beam steering system would be small and lightweight so that it can be mounted on a vehicle or a satellite, or integrated on a handheld device, such as a smartphone [4]. However, the state-of-the-art optical beam steering systems typically comprise mechanical assemblies of moving parts and bulk optic components [5]. Optical phased arrays (OPAs) have gained significant interest as a static, non-mechanical beam-steering devices [6–9]. OPAs can be integrated on chip to achieve beam steering by controlling the phase of the light emitted by the antennas forming the array. Optical antennas are the fundamental elements in the on-chip OPAs [10–13], with antenna efficiency, the aperture size and the far-field radiation angle being the key parameters determining the OPA performance. Surface grating couplers have been extensively used in silicon-based planar waveguides for fiber-chip coupling and wafer-scale testing [14–21]. However, for the maximum overlap with optical fiber mode and high coupling efficiency, grating couplers are usually 10-15 micrometers long. Substantially more compact antenna designs are required for dense OPAs with a wide beam steering range [22–27]. An appealing solution to reduce the antenna dimension while maintaining a high efficiency is to enhance the grating strength by using the 300 nm SOI platform [11,28–30]. We had recently demonstrated a high diffraction efficiency exceeding 89% and a compact footprint of 7.6 µm × 4.5 µm on a 300 nm SOI [31]. While this constitutes an important advance in terms of antenna footprint and efficiency, a remaining outstanding challenge is the antenna's limited far field radiation angle, restricting the OPA's steering range.

In this work we demonstrate, for the first time, a novel design of an ultra-compact silicon grating antenna for off-chip light emission, with an unprecedented performance combining a wide far-field beam width, a very compact footprint, and a high diffraction efficiency. This is achieved with a single-etch grating structure designed on a 300-nm-thick SOI platform. The design is performed using 2D Finite-Difference Time-Domain (FDTD) simulations, followed by validation using 3D FDTD simulations. The design methodology is described in section 2. The simulation results are presented in section 3, and the conclusions are summarized in section 4.

## 2. Antenna far-field broadening

The far-field distribution generated by an optical antenna is related to the near field through the Fraunhofer transformation [32]:

$$U(x,y,z) \propto \iint_A E(x',y') e^{-i\left(\frac{k(x'x+y'y)}{z}\right)} dx'dy' \quad (1)$$

Where $E(x',y')$ is the complex amplitude at the aperture A located at $x'y'$ plane, and k is the wave vector. The shape and size of the antenna aperture directly affects the far-field distribution. While large beam widths in the far-field can be obtained by using small apertures, reducing the aperture size limits the antenna's efficiency.

Here we propose a new strategy to circumvent this fundamental limit, to leverage the near-field phase engineering. Instead of adjusting the antenna aperture, the key idea is to broaden the far-field by judiciously controlling the phase of the complex near field $E(x',y')$. By introducing a specific near-field phase factor in the Fraunhofer transformation, the far-field beam width can be widened without shortening the antenna. This strategy allows us to decouple the minimum antenna length (requirement for a given scattering efficiency) from the far-field beam width.

The near field phase can be altered by longitudinally chirping the antenna grating period. However, as we show later in section 3.2, for short antennas with strong radiation strength the chirping effect is ineffective because the antenna comprises only a small number of periods (e.g., 3 - 5) and their respective contributions to the scattered field are rapidly decreasing along the antenna. If a conventional longitudinal chirp is implemented in such short antennas, the contribution of the last periods to the scattered field is substantially smaller compared to the first period. This constraint fundamentally limits the effectiveness of the chirping effect. Instead of chirping the grating longitudinally, here we use, for the first time, transversally interleaved chirped nanostructures to control the near-field phase. With transverse chirping, the different chirp periods equally contribute to the scattered field, enabling effective phase engineering in very short (few-period) gratings and yielding a substantially broadened far-field radiation compared to the conventional antennas.

## 3. Design and simulation results
### 3.1 Periodic grating antenna

As a reference case of antenna with high scattering efficiency [33], we consider a single-etch, periodic grating in a standard SOI platform with a 300-nm-thick silicon waveguide core, 1-μm buried oxide (BOX) and 2-μm oxide cladding. The grating comprises three periods, each including two fully etched sections with lengths $L_1$ and $L_3$ and two un-etched sections with length $L_2$ and $L_4$, as schematically shown in Fig. 1.

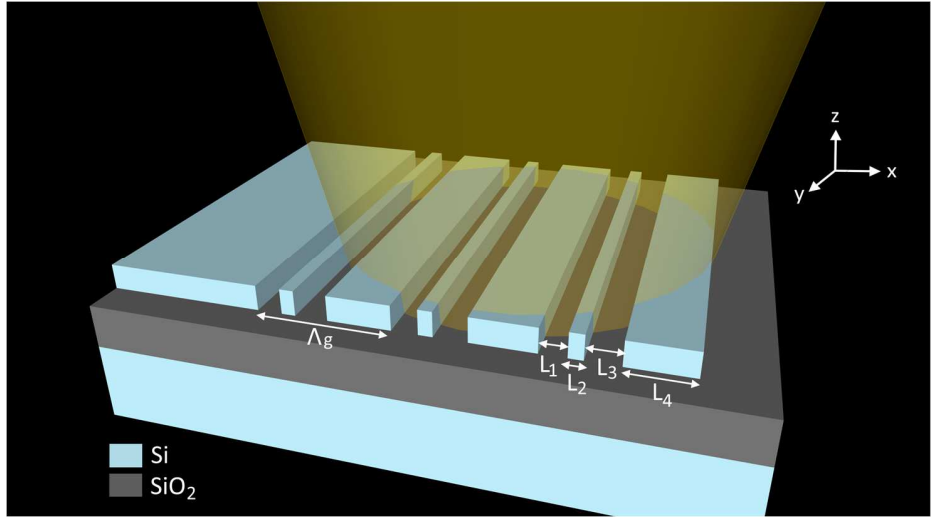

**Figure 1.** Schematic of the single-etch, un-chirped antenna with grating period $\Lambda_g$ and structural parameters [$L_1$, $L_2$, $L_3$, $L_4$]. A transverse electric (TE) waveguide mode is incident from the left.

The four geometrical parameters $L_1$-$L_4$ are optimized using a genetic algorithm combined with 2D Finite-Difference Time-Domain (FDTD) simulations in order to achieve the maximum upward diffraction efficiency while the diffraction angle is allowed to vary as a free parameter. The detailed optimization procedure can be found in our previous work [31]. The optimized parameters of the nanostructure forming the grating period of $\Lambda_g$=733 nm are $L_1$=109 nm, $L_2$=48 nm, $L_3$=189 nm, and $L_4$=387 nm, while the full antenna length is 2.2 μm. Rigorous 3D FDTD simulations are then used to validate the results of the 2D analysis. The antenna width for this structure is chosen as 1.75 μm, to yield a compact design required for dense antenna arrays. The upward diffraction efficiency at λ = 1550 nm obtained from 2D and 3D simulations is 60% and 55%, respectively, while the grating reflectivity computed as the fraction of back-reflected power coupled to the counter propagating TE mode of the input waveguide is -20 dB (2D) and -12 dB (3D). It should be noted that shortening the grating further and using two grating periods instead of three would significantly reduce the diffraction efficiency of the grating, by 11%, while adding a fourth period would only increase efficiency by 2%. Fig. 2 shows the far-field radiation pattern for the reference antenna design, where θ and ϕ represent the polar and azimuthal angles, respectively. The peak diffraction angle is 7° from the vertical, and the full width and half maximum (FWHM) of the far-field intensity along the longitudinal (x) and transverse (y) directions are 36° and 50° at 1550 nm, respectively.

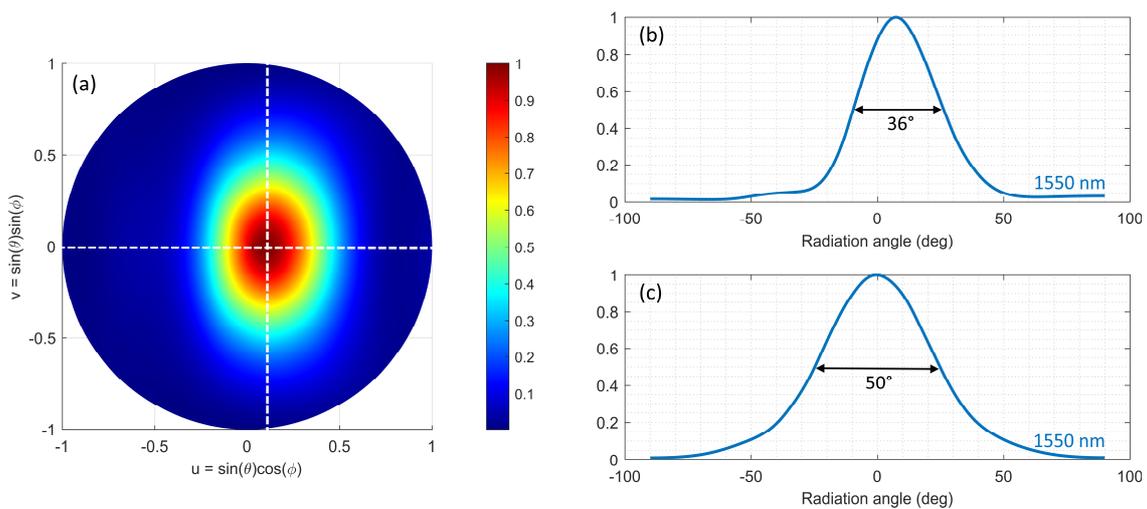

**Figure 2.** (a) The far-field radiation pattern of the reference antenna at a wavelength of 1550 nm. The far-field intensity cuts along the x (b) and y axis (c), as marked by the dashed white lines in (a).

### 3.2 Longitudinally-chirped antenna

In order to evaluate the effect of longitudinal chirping on the far-field beam width, we design an antenna with the structure described in section 3.1, but considering different periods for the three unit cells. The goal is to engineer the near-field phase profile of the diffracted beam in order to increase the divergence and obtain a broader beam width in the far-field. Simulating a range of grating periods showed that $\Lambda_{g1} < 600$ nm results in significant reduction in upward diffraction efficiency. The first cell period was hence designed to provide a large negative diffraction angle of -21°. Cell parameters were optimized as described in the previous section, resulting in $L_1 = 133$ nm, $L_2 = 61$ nm, $L_3 = 190$ nm, and $L_4 = 216$ nm (period $\Lambda_{g1} = 600$ nm). For the second cell, we use the optimized structural parameters from the reference design of section 3.1 ($\Lambda_{g2} = 733$ nm, diffraction angle of 7°). For the last cell period we choose a positive diffraction angle of 9°, which still avoids the second-order diffraction. The optimized parameters are $L_1 = 109$ nm, $L_2 = 41$ nm, $L_3 = 300$ nm, and $L_4 = 350$ nm, i,e, cell period of $\Lambda_{g3} = 800$ nm. The width of the antenna is 1.75 μm; that is, the same as the reference structure.

3D FDTD was used to simulate the behavior of the longitudinally-chirped grating. We obtained a FWHM of the far-field beam of 38° and 50.2° in the x and y directions, respectively. Diffraction efficiency was reduced to 45% compared to reference design, while maintaining comparable back-reflection (-10 dB). Despite the large chirping from $\Lambda_{g1} = 600$ nm to $\Lambda_{g3} = 800$ nm, the beam width in the longitudinal direction increased only marginally by 2°, indicating the limited effectiveness of longitudinal chirping to broaden the far-field of short antennas.

### 3.3 Transversally-interleaved chirped antenna

As described in section 2, we propose to overcome the ineffectiveness of longitudinal chirping in few-period antennas using transversal chirping.

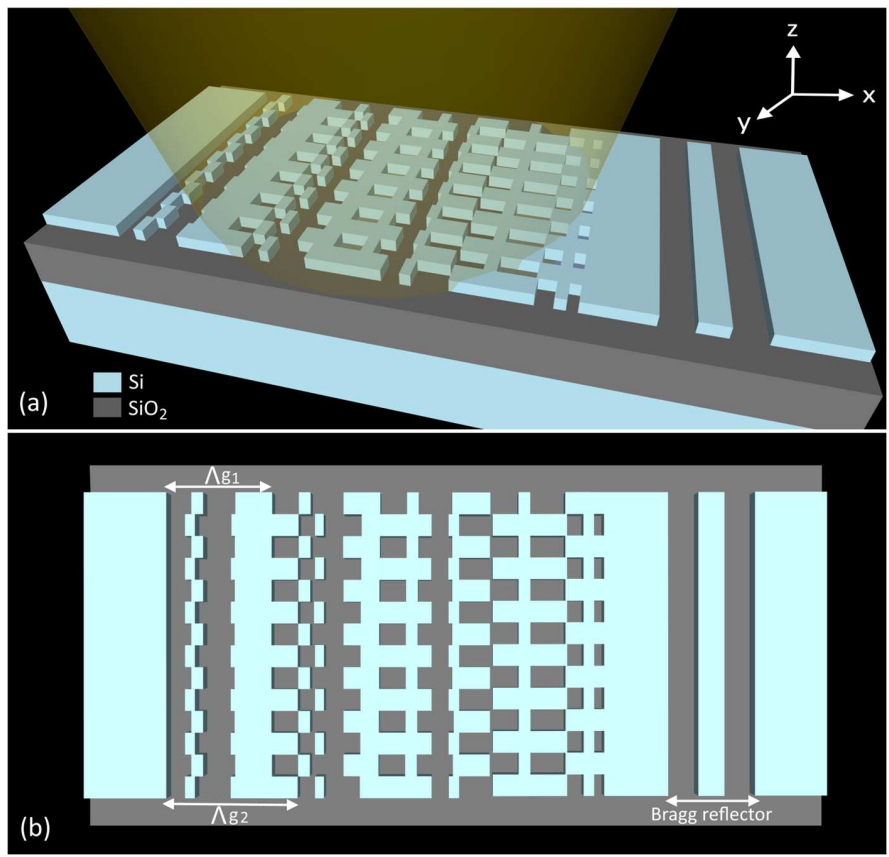

**Figure 3.** (a) Transversally interleaved antenna schematics. The incident light propagates along the x direction, being diffracted of-chip by the antenna nanostructure. The residual power propagating in the waveguide plane is re-injected to the antenna by a terminal Bragg reflector. (b) The antenna top view (xy plane).

Instead of chirping the grating period along the propagation direction, the antenna is implemented by transversally interleaving two gratings with different longitudinal periods, as schematically shown in Fig. 3. Unlike traditional chirped grating, the transverse chirping allows for the different chirp periods equally contribute to the scattered field because the scatterers are arranged in parallel rather than sequentially.

We used two of the optimized grating designs described in section 3.2, the first with diffraction angle of -21° (longitudinal period 600 nm) and the second with diffraction angle of 7° (longitudinal period 733 nm). The two gratings are transversally interleaved fourteen times with a subwavelength pitch of 125 nm, resulting in a total antenna width of 1.75 μm. Each grating has four periods along the longitudinal direction. We find that transversal chirping increases the far-field beam width in the longitudinal direction (as estimated by 3D FDTD simulations) to 43°, an increase of 7° compared to the reference grating described in section 3.1.

A Bragg reflector is placed at the end of the grating to re-circulate the part of the light left un-diffracted at the end of the grating, back into the antenna, hence improving its efficiency. The reflector comprises two periods of silicon segment with a width of 148 nm separated by 152 nm gaps. The calculated Bragg mirror reflectivity is 95% at a wavelength of 1550 nm. The distance to the reflector from the antenna is judiciously optimized to obtain constructive interference between the forward propagating mode diffracted by the grating and the backward propagating mode diffracted upon reflection. The maximum upward diffraction efficiency of 51% is obtained for the reflector-grating separation of 300 nm, while only 5% of the power remains un-diffracted after the first pass through the antenna and the Bragg mirror. The calculated efficiency is comparable with that of the reference periodic grating, showing that we almost fully compensate the weakening of the grating strengths due to the chirping effect. Furthermore, the use of the Bragg reflector allows to further widen the far-field beam width in the longitudinal direction, as the result of the reversal of the diffraction angle. A gain of an additional 8° is observed in longitudinal beam width, yielding a wide far-field beam of 51°×47° centered near -12°. In total, the beam width in the longitudinal direction is about 15° larger than the reference case while in transversal direction it remains approximately unchanged.
It is important to note that the increased beam divergence is obtained despite the aperture of the interleaved antenna (3 μm × 1.75 μm) being larger than that of the reference design (2.2 μm × 1.75 μm). This confirms that it is possible to circumvent the limitation on far-field beam size for a given antenna aperture by effectively engineering the near-field phase profile of the diffracted light.

Since a relatively large fraction of light (~30%) is diffracted downward, this light can be re-directed upward to further enhance the efficiency of the antenna. This can be achieved by using a bottom reflector based on a double-SOI substrate, similar to that presented in [34]. The bottom Bragg reflector is designed by optimizing the thickness of different layers of the double SOI structure. With a silicon thickness of 110 nm sandwiched between two silicon dioxide layers with thicknesses of 500 nm and 274 nm, a reflectivity of about 71% is obtained at the central wavelength of 1550 nm. By incorporating the bottom mirror in our antenna, a diffraction efficiency of 82% and the far-field beam width of 52°×62° are obtained as shown in Fig. 4. The downward diffracted power is reduced to 8% and the back-reflectivity of the antenna is -16 dB, while the peak diffraction angle remains almost unchanged at -13°.

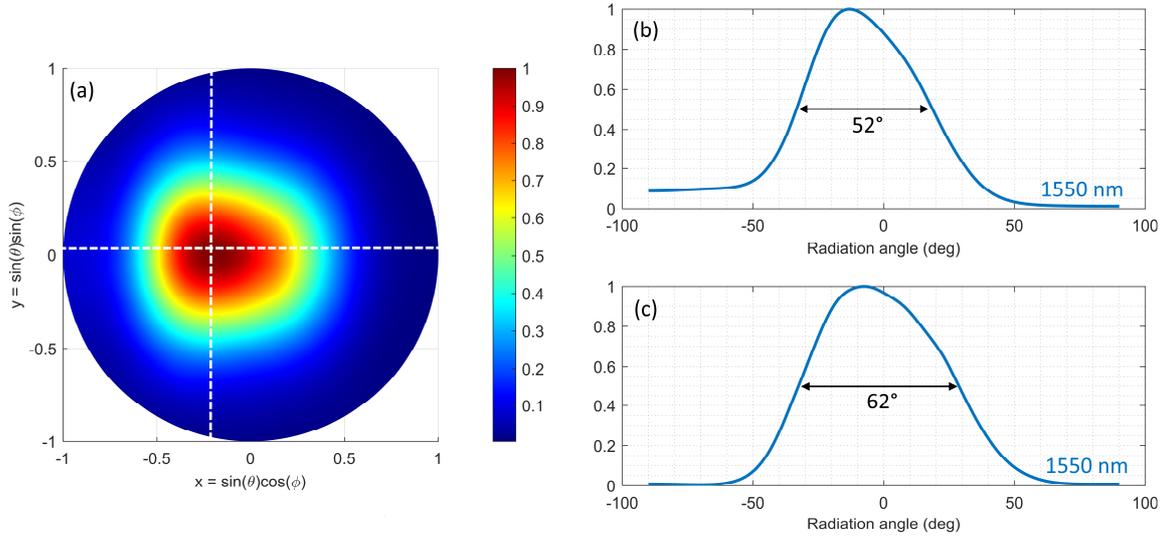

**Figure 4.** (a) The far-field intensity of the optimized antenna design at 1550 nm wavelength. (b) Far-field intensity distribution along the x and (c) y axis, as marked by the dashed white lines in (a).

To compare our antenna with the state-of-the-art results, we note that in [35] a dual-input antenna based on two waveguides placed on opposite sides of the antenna was used to increase the beam width. Their reported simulated angular range for an effective antenna aperture of 2.5 μm$^2$ is limited to 40°×15°, much smaller than our exhibited 52°×62°. This confirms the effectiveness of our transverse chirping strategy that substantially broadens the antenna fa-field.

## 4. Conclusion

In this paper, we demonstrate a new strategy to circumvent the fundamental limitation of the antenna beam width by near-field phase engineering. This is achieved by using transversally interleaved subwavelength grating nanostructure. Our optimized antenna design achieves a FWHM of 52°×62°, a high diffraction efficiency of 82% and an ultra-compact footprint of 3.15 μm × 1.75 μm. Based on these results, we expect that exploration of near-field phase engineered nanostructures and integrated photonic components using transverse interleaved subwavelength gratings will lead to a new research direction in the development of integrated photonics antennas for a wide range of applications, including free-space optical interconnects and on-chip optical phased arrays for lidar systems.

## References

1. Doylend, J. K. *et al.* Two-dimensional free-space beam steering with an optical phased array on silicon-on-insulator. *Opt. Express* **19**, 21595 (2011).
2. Zhang, Y., Ling, Y. C., Zhang, K. & Ben Yoo, S. J. Sub-Wavelength-Pitch Silicon-Photonic Optical Phased Array for Large Field-AF-Regard Coherent Optical Beam Steering. *Eur. Conf. Opt. Commun. ECOC* **2018-Sept**, 1929–1940 (2018).
3. Guo, Y. *et al.* Integrated optical phased arrays for beam forming and steering. *Appl. Sci.* **11**, (2021).
4. Heck, M. J. R. Highly integrated optical phased arrays: Photonic integrated circuits for optical beam shaping and beam steering. *Nanophotonics* **6**, 93–107 (2017).
5. Kim, T. *et al.* A single-chip optical phased array in a wafer-scale silicon photonics/CMOS 3D-integration platform. *IEEE J. Solid-State Circuits* **54**, 3061–3074 (2019).
6. He, J., Dong, T. & Xu, Y. Review of photonic integrated optical phased arrays for space optical communication. *IEEE Access* **8**, 188284–188298 (2020).
7. Abediasl, H. & Hashemi, H. Monolithic optical phased-array transceiver in a standard SOI CMOS process. *Opt. Express* **23**, 6509–6519 (2015).

8. Heck, M. J. R. & Access, O. Review article Highly integrated optical phased arrays : photonic integrated circuits for optical beam shaping and beam steering. **6**, 93–107 (2017).
9. Sun, J., Timurdogan, E., Yaacobi, A., Hosseini, E. S. & Watts, M. R. Large-scale nanophotonic phased array. *Nature* **493**, 195–199 (2013).
10. Pita, J. L., Aldaya, I., Dainese, P., Hernandez-Figueroa, H. E. & Gabrielli, L. H. Design of a compact CMOS-compatible photonic antenna by topological optimization. *Opt. Express* **26**, 2435–2442 (2018).
11. Melati, D. *et al.* Design of compact and efficient silicon photonic micro antennas with perfectly vertical emission. *IEEE J. Sel. Top. Quantum Electron.* **27**, 1–10 (2020).
12. Fatemi, R., Khachaturian, A. & Hajimiri, A. A nonuniform sparse 2-D large-FOV optical phased array with a low-power PWM drive. *IEEE J. Solid-State Circuits* **54**, 1200–1215 (2019).
13. Zhang, X., Kwon, K., Henriksson, J., Luo, J. & Wu, M. C. A large-scale microelectromechanical-systems-based silicon photonics LiDAR. *Nature* **603**, 253–258 (2022).
14. Vermeulen, D. *et al.* High-efficiency fiber-to-chip grating couplers realized using an advanced CMOS-compatible Silicon-On-Insulator platform. *Opt. Express* **18**, 18278–18283 (2010).
15. Melati, D. *et al.* Mapping the global design space of nanophotonic components using machine learning pattern recognition. *Nat. Commun.* **10**, 1–9 (2019).
16. Watanabe, T., Ayata, M., Koch, U., Fedoryshyn, Y. & Leuthold, J. Perpendicular grating coupler based on a blazed antiback-reflection structure. *J. Light. Technol.* **35**, 4663–4669 (2017).
17. Benedikovic, D. *et al.* High-directionality fiber-chip grating coupler with interleaved trenches and subwavelength index-matching structure. *Opt. Lett.* **40**, 4190–4193 (2015).
18. Cheben, P., Halir, R., Schmid, J. H., Atwater, H. A. & Smith, D. R. Subwavelength integrated photonics. *Nature* **560**, 565–572 (2018).
19. Chen, X., Thomson, D. J., Crudginton, L., Khokhar, A. Z. & Reed, G. T. Dual-etch apodised grating couplers for efficient fibre-chip coupling near 1310 nm wavelength. *Opt. Express* **25**, 17864–17871 (2017).
20. Zou, J., Yu, Y. & Zhang, X. Single step etched two dimensional grating coupler based on the SOI platform. *Opt. Express* **23**, 32490–32495 (2015).
21. Tong, Y., Zhou, W. & Ki Tsang, H. Efficient perfectly vertical grating coupler for multi-core fibers fabricated with 193 nm DUV lithography. **43**, 5709–5712 (2018).
22. Alonso-Ramos, C. *et al.* Fiber-chip grating coupler based on interleaved trenches with directionality exceeding 95%. *Opt. Lett.* **39**, 5351–5354 (2014).
23. Halir, R. *et al.* Continuously apodized fiber-to-chip surface grating coupler with refractive index engineered subwavelength structure. *Opt. Lett.* **35**, 3243–3245 (2010).
24. Bozzola, A., Carroll, L., Gerace, D., Cristiani, I. & Andreani, L. C. Optimising apodized grating couplers in a pure SOI platform to −0.5 dB coupling efficiency. *Opt. Express* **23**, 16289–16304 (2015).
25. Chen, H. Y. & Yang, K. C. Design of a high-efficiency grating coupler based on a silicon nitride overlay for silicon-on-insulator waveguides. *Appl. Opt.* **49**, 6455–6462 (2010).
26. Wang, Y. *et al.* Design of broadband subwavelength grating couplers with low back reflection. *Opt. Lett.* **40**, 4647–4650 (2015).
27. Kamandar Dezfouli, M. *et al.* Perfectly vertical surface grating couplers using subwavelength engineering for increased feature sizes. *Opt. Lett.* **45**, 3701–3704 (2020).
28. Benedikovic, D. *et al.* Sub-decibel silicon grating couplers based on L-shaped waveguides and engineered subwavelength metamaterials. *Opt. Express* **27**, 26239–26250 (2019).
29. Benedikovic, D. *et al.* High-efficiency single etch step apodized surface grating coupler using subwavelength structure. *Laser Photon. Rev.* **8**, L93–L97 (2014).
30. Passoni, M., Gerace, D., Carroll, L. & Andreani, L. C. Grating couplers in silicon-on-insulator: The role of photonic guided resonances on lineshape and bandwidth. *Appl. Phys. Lett.* **110**, 41107–41111 (2017).
31. Khajavi, S. *et al.* Compact and highly-efficient broadband surface grating antenna on a silicon platform. *Opt. Express* **29**, 7003–7014 (2021).
32. Hecht, E. & Black, A. Fraunhofer diffraction. *Opt. Reading, MA Addison Wesley Longman, Inc* 452–485 (2002).
33. Benedikovic, D. *et al.* Single-etch subwavelength engineered fiber-chip grating couplers for 13 μm datacom wavelength band. *Optics Express* vol. 24 12893–12904 (2016).
34. Baudot, C. *et al.* Low cost 300mm double-SOI substrate for low insertion loss 1D & 2D grating couplers. *IEEE Int. Conf. Gr. IV Photonics GFP* **1**, 137–138 (2014).
35. Fatemi, R., Khial, P. P., Khachaturian, A. & Hajimiri, A. Breaking FOV-Aperture Trade-Off With Multi-Mode Nano-Photonic Antennas. *IEEE J. Sel. Top. Quantum Electron.* **27**, 1–14 (2020).

## Additional information
**Competing interests:** The authors declare no competing interests.

## Funding
National Research Council Canada: (HTSN 624), (CSTIP), Technology and Innovation Program; Natural Sciences and Engineering Research Council of Canada.